\documentclass[11pt]{article}
\usepackage{amsmath}
\usepackage{amssymb}
\usepackage{amsthm,amsxtra}
\usepackage{epsf}
\usepackage{eepic}
\newlength{\mytopmargin}
\newlength{\myleftmargin}
\newcommand{\zz}{\mathbb Z}

\begin{document}
\vspace{2cm}
\noindent
\begin{center}{\Large \bf Spacing distributions in random
matrix ensembles}
\end{center}
\vspace{5mm}

\begin{center}
P.J.~Forrester
\end{center}

\vspace{.2cm}

\noindent
Department of Mathematics and Statistics, University of
Melbourne, Victoria 3010, Australia; \\

\section{Introduction}
\subsection{Motivation and definitions}
The topic of spacing distributions in random matrix ensembles is almost
as old as the introduction of random matrix theory into nuclear physics.
Both events can be traced back to Wigner in the mid 1950's \cite{Wi55,
Wi57}. Thus Wigner introduced the model of a large real symmetric random
matrix, in which the upper triangular elements are independently
distributed with zero mean and constant variance, for purposes of
reproducing the statistical properties of the highly excited energy levels
of heavy nuclei. This was motivated by the gathering of experimental data
on the spectrum of isotopes such as ${}^{238}$U at energy levels beyond
neutron threshold. Wigner hypothesized that the statistical properties of
the highly excited states of complex nuclei would be the same as those of the
eigenvalues of large random real symmetric matrices.
For the random matrix model to be of use at a 
quantitative level, it was necessary to deduce analytic forms of 
statistics of the eigenvalues which could be compared against statistics
determined from experimental data.

What are natural statistics for a sequence of energy levels, and can these
statistics be computed for the random matrix model? Regarding the first
question, let us think of the sequence as a point process on the line, and
suppose for simplicity that the density of points is uniform and has been
normalized to unity. For any point process in one dimension a fundamental
quantity is the probability density function for the event that given
there is a point at the origin, there is a point in the interval
$[s,s+ds]$, and further there are $n$ points somewhere in between 
these points and thus in the interval
$(0,s)$. Let us denote the probability density function
by $p(n;s)$. In the language of energy levels,
this is the spacing distribution between levels $n$ apart. 

Another fundamental statistical quantity is the $k$-point distribution
function $\rho_{(k)}(x_1,\dots,x_k)$. This can be defined recursively, starting
with $\rho_{(1)}(x)$, by the requirement that
\begin{equation}\label{r2.a}
\rho_{(k)}(x_1,\dots,x_k)/\rho_{(k-1)}(x_1,\dots,x_{k-1})
\end{equation}
is equal to the density of points at $x_k$, given there are points at
$x_1,\dots, x_{k-1}$. One sees from the definitions that
\begin{equation}\label{r2}
{\rho_{(2)}(0,s) \over \rho_{(1)}(0)} =
\sum_{n=0}^\infty p(n;s).
\end{equation}
From empirical data of a long energy level sequence, the quantity $p(n;s)$
for small values of $n$ at least is readily estimated (the statistical
uncertainty gets worse as $n$ increases). Use of (\ref{r2})
then allows for an estimation of $\rho_{(2)}(0;s)$.

We thus seek the theoretical determination of $p(n;s)$ for matrix
ensembles.

\subsection{Spacing between primes}
Before taking up the problem of determining $p(n;s)$ for matrix ensembles, 
which is the theme of these lectures, let
us digress a little and follow the line of introduction to spacing 
distributions given by Porter in the review he wrote as part of the book
\cite{Po65}, which collected together the major papers written in the field
up to 1965. Porter's introduction is particularly relevant to the theme 
of the present school because it uses the prime numbers as an example of
a deterministic sequence which, like energy levels of heavy nuclei, 
nevertheless exhibit pronounced stochastic features.

It turns out the spacing distributions between primes relate to perhaps
the simplest example of a point process. This is when the probability that
there is a point in the interval $[s,s+ds]$ is equal to $ds$, independent
of the location of the other points. This generates the so called Poisson
process with unit density, or in the language of statistical mechanics, a
perfect gas. By definition of the process the ratio (\ref{r2.a}) is unity
for all $k$ and thus
\begin{equation}\label{r3b}
\rho_{(k)}(x_1,\dots,x_k)=1.
\end{equation}
To compute $p(n;s)$, we think of the Poisson process as the $N \to \infty$
limit of a process in which each unit interval on the line is broken up into
$N$ equal sub-intervals, with the probability of there being a particle in
any one of the subintervals equal to $1/N$. Thus
\begin{equation}\label{r3}
p(s;n) = \lim_{N \to \infty} ( 1 - {1 \over N} )^{sN-n} N^{-n}
\Big ( {sN \atop n} \Big ) = {s^n \over n!} e^{-s}.
\end{equation}
In the first equality of (\ref{r3}), the first factor is the probability that
$sN-n$ subintervals do not contain a particle, the second factor is the
probability that $n$ subintervals do contain a particle, while the final factor
is the number of ways of choosing $n$ occupied sites amongst $sN$ sites in
total. The probability density in the final equality of (\ref{r3}) is the
Poisson distribution. Substituting (\ref{r3}) in (\ref{r2}) gives
$\rho_{(2)}(0,x)=1$, as required by (\ref{r3b}).

The distribution (\ref{r3}) ties in with prime numbers through Kram\'er's
model (see the lectures by Heath-Brown in the present volume). In
this approximation, statistically the primes are regarded as forming a
Poisson process on the positive integer lattice. The probability of
occupation of the $N$th site is taken to equal $1/\log N$, so as to be
consistent with the prime number theorem. Kram\'er's model predicts that as
an approximation
\begin{equation}\label{r4}
p^{(N)}(n;s) = {s^n \over n!} e^{-s}, \qquad s = t/\log N
\end{equation}
where $p^{(N)}(n;s)$ refers to the probability that for primes $p$ in the
neighbourhood of a prime $N$, there is a prime at $p+t$, and furthermore
there are exactly $n$ primes between $p$ and $p+t$.

\begin{figure}
\epsfxsize=12cm
\centerline{\epsfbox{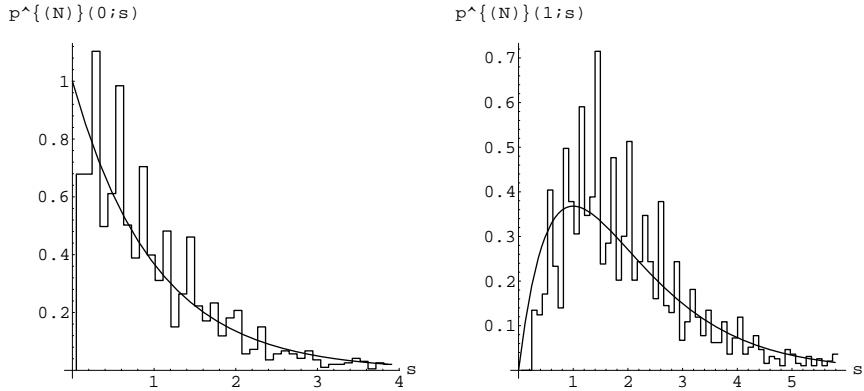}}
\caption{\label{af.1} Distribution of the spacing $t$ between primes (leftmost
graph) and the spacing $t$ between every second prime for $2,000$ consecutive
primes starting with $N=10^9+7$. The distributions are given in units of
$s=t/\log N$. The smooth curves are the Poisson distributions $p(0;s) =
e^{-s}$ and $p(1;s)= se^{-s}$. }
\end{figure}

To compare the prediction (\ref{r4}) against empirical data, we choose a
value of $N$, say $10^9$, and for the subsequent $M$ primes (say 
$M=2,000$) record the distance to the following prime (in relation to
$p^{(N)}(1;s)$) and the distance to the second biggest prime after that
(in relation to $p^{(N)}(s;1)$). We form a histogram, with the scale on the
horizontal axis measured in units of $s=t/\log N$, where $t$ is the actual
spacing. The natural units for $t$ are multiples of 2, and this provides
a width for the bars of the histogram. We see from Figure \ref{af.1}
that the general trend of the histograms do indeed follow the
respective Poisson distributions.

\vspace{.5cm}

\subsection{Empirical determination of spacing distributions for matrix
ensembles}
Wigner's interest was in the statistical properties of the eigenvalues
of large real symmetric random matrices. More particularly, he sought the
statistical properties of the eigenvalues in what may be termed the bulk of
the spectrum (as opposed to the edge of the spectrum \cite{Fo93a}). The
eigenvalues in this region are characterized by having a uniform
density, which after rescaling (referred to as `unfolding') may be taken as
unity. In distinction to the situation with the sequence of primes, for
random matrices it is not necessary to study the statistical properties of
a large sequence of (unfolded) eigenvalues from a single matrix. Rather
the spacing distributions with respect to the middle eigenvalue (this is
the eigenvalue most in the bulk) in multiple samples from the class
of random matrices in question can be listed, and then this list used to 
create a histogram. Moreover, to approximate large matrix size
behaviour, it is only necessary to consider quite small matrix sizes,
say $13 \times 13$.

\begin{figure}
\epsfxsize=12cm
\centerline{\epsfbox{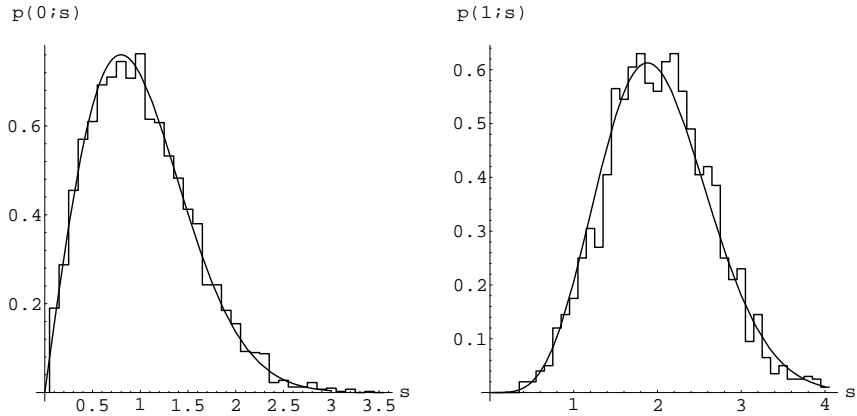}}
\caption{\label{af.2} Plot of the distribution of the unfolded spacing
between the 6th and 7th, and 7th and 8th eigenvalues
(pooled together) for 2,000 samples from the $13\times 13$ GUE
eigenvalue distribution. The smooth curve is the Wigner surmise
(\ref{ws}). The rightmost graph is the distribution between the 6th and
8th eigenvalues in the same setting, while in the smooth curve in this
case is $(1/2) p_4(0;s/2)$ with $p_4$ given by (\ref{2.10b}).
}
\end{figure}

In Figure \ref{af.2} we have plotted the empirical determination of
$p(0;s)$ and $p(1;s)$ obtained from lists of eigenvalue spacings for
realizations of the so called GOE (Gaussian orthogonal ensemble) eigenvalue
distribution. 
As we know from the lectures of Fyodorov in this volume, the
GOE consists of
real symmetric random matrices, with each diagonal element chosen from the
normal distribution N$[0,1]$, and each (strictly) upper triangular element
chosen from the normal distribution N$[0,1/\sqrt{2}]$. For such matrices,
it is well known that to leading order in the matrix rank $N$, the
eigenvalue density is given by the Wigner semi-circle law
$$
\rho_{(1)}(x) = {\sqrt{2N} \over \pi} \sqrt{ 1 - {x^2 \over 2 N} }.
$$
Multiplying the eigenvalues at point $x$ by this factor allows us to unfold
the sequence giving a mean eigenvalue spacing of unity.

A less well known, and much more recent result relating to GOE matrices is
that their spectrum can be realized without having to diagonalize a 
matrix \cite{DE02} (see also \cite{FR02b}).   Thus one has that the roots
of the random polynomial $P_N(\lambda)$, defined recursively by the
stochastic three term recurrence
\begin{equation}\label{r7}
P_k(\lambda) = (\lambda - a_k) P_{k-1}(\lambda) - b_{k-1}^2 P_{k-2}(\lambda)
\end{equation}
where
$$
a_k \: \sim \: {\rm N}[0,1], \qquad
b_k^2 \: \sim \: {\rm Gamma}[k/2,1],
$$
have the same distribution as the eigenvalues of GOE matrices
(the notation Gamma$[s,\sigma]$ denotes the gamma distribution with
density proportional to $x^{s-1} e^{-x/\sigma}$). Generating
such polynomials and finding their zeros then provides us with a
sequence 
distributed as for GOE
eigenvalues, from which we have determined $p(0;s)$ and
$p(1;s)$.

\vspace{.5cm}

\section{Eigenvalue product formulas for gap probabilities}
\setcounter{equation}{0}
\subsection{Theory relating to $p(n;s)$}
Consider a point process consisting of a total of $N$ points. Let the joint
probability density function of the $N$ points be denoted $p(x_1,\dots,
x_N)$. A quantity closely related to the spacing distribution $p(0;s)$
is the gap probability
\begin{equation}
E^{\rm bulk}(0;s) := \lim_{N \to \infty} a_N^N \int_{\bar{I}} dx_1
\cdots \int_{\bar{I}} dx_N \, p(a_N x_1,\dots, a_N x_N)
\end{equation}
where $\bar{I} = (-\infty,\infty) - (-s/2,s/2)$ and $a_N$ is the leading large
$N$ form of the local density at the origin (and thus the unfolding factor).
Thus it is easy to see that
\begin{equation}\label{pE}
p(0;s) = {d^2 \over d s^2} E^{\rm bulk}(0;s).
\end{equation}

More generally we can define
\begin{equation}
E^{\rm bulk}(n;s) := \lim_{N \to \infty} \Big ( {N \atop n} \Big )
a_N^N \int_{-s/2}^{s/2} dx_1 \cdots \int_{-s/2}^{s/2} dx_n
\int_{\bar{I}} dx_{n+1} \cdots \int_{\bar{I}} dx_N \,
p(a_N x_1,\dots, a_N x_N).
\end{equation}
These quantities can be calculated from the generating function
\begin{equation}\label{ap}
E^{\rm bulk}(s;\xi) := \lim_{N \to \infty} a_N^N 
\int_{-\infty}^\infty dx_1 \cdots \int_{-\infty}^\infty dx_N \,
\prod_{l=1}^N (1 - \xi \chi_{(-s/2,s/2)}^{(l)} )
p(a_N x_1,\dots, a_N x_N),
\end{equation}
where $\chi_J^{(l)} = 1$ for $x^{(l)} \in J$ and 
$\chi_J^{(l)} = 0$ otherwise,
according to the formula
\begin{equation}\label{2.5x}
E^{\rm bulk}(n;s) = {(-1)^n \over n!} {\partial^n \over \partial \xi^n }
E^{\rm bulk}(s;\xi) \Big |_{\xi = 1}.
\end{equation}
It follows from the definitions that
\begin{equation}
p(n;s) = {d^2 \over ds^2} E^{\rm bulk}(n;s) + 2 p(n-1;s) -
p(n-2;s),
\end{equation}
or equivalently
\begin{equation}
p(n;s) = {d^2 \over ds^2} \sum_{j=0}^n (n-j+1) E^{\rm bulk}(j;s).
\end{equation}
Hence knowledge of $\{E^{\rm bulk}(j;s)\}_{j=0,\dots,n}$ is sufficient for
the calculation of $p(n;s)$.

It is possible to relate (\ref{ap}) to the $k$-point distribution
functions. In the finite system the latter are given by
\begin{equation}\label{ap1}
\rho_{(k)}^{(N)}(x_1,\dots,x_k) = {N! \over (N-k)!}
\int_{-\infty}^\infty dx_{k+1} \cdots \int_{-\infty}^\infty dx_N \,
p(x_1,\dots,x_N).
\end{equation}
With
$$
\rho_{(k)}^{\rm bulk}(x_1,\dots,x_k) := \lim_{N \to \infty} a_N^k
\rho_{(k)}^{(N)}(a_N x_1,\dots,a_N x_k),
$$
by expanding (\ref{ap}) in a power series in $\xi$ and making use of
(\ref{ap1}) we see that
\begin{equation}\label{ap2}
E^{\rm bulk}(s;\xi) = 1 +
\sum_{k=1}^\infty {(-\xi)^k \over k!}
\int_{-s/2}^{s/2} dx_1 \cdots \int_{-s/2}^{s/2} dx_k \,
\rho_{(k)}^{\rm bulk}(x_1,\dots,x_k).
\end{equation}
For the limiting process to be rigorously justified, because $[-s/2,s/2]$
is a compact interval, it is sufficient that 
$\rho_{(k)}^{\rm bulk}(x_1,\dots,x_k)$ be bounded by $M^k$ for some 
$M > 0$.

With these basic formulas established, we will now proceed to survey some
of the main results relating to spacing distributions in the bulk of the
various matrix ensembles (orthogonal, unitary and symplectic symmetry
classes).

\subsection{Wigner surmise}
For the Poisson process we have seen that $p(0;s) = e^{-s}$. Thus in this
case the spacing distribution is actually maximum at zero separation
between the points. The opposite feature is expected for $p(0;s)$ in relation 
to the eigenvalues of random real symmetric matrices, as can be seen by
examining the $2 \times 2$ case of matrices of the form
$$
A = \left [ \begin{array}{cc} a & b \\ b & c \end{array}
\right ].
$$
This matrix is diagonalized by the decomposition
$A = R {\rm diag}[\lambda_+,\lambda_-] R^T$ where
$$
R = \left [ \begin{array}{cc} \cos \theta & - \sin \theta \\
\sin \theta &  \cos \theta  \end{array} \right ].
$$
Expressing $a,b,c$ in terms of $\lambda_+,\lambda_-, \theta$ it is simple to
show
\begin{equation}\label{2.10'}
da db dc = |\lambda_+ - \lambda_-| d \lambda_+ d \lambda_- d \theta.
\end{equation}
Thus for small separation $s:= |\lambda_+ - \lambda_-|$ the probability density
function vanishes linearly.

Let $\mu(s)$ denote the small $s$ behaviour of $p(0;s)$. We have seen that
for the Poisson process $\mu(s)=1$, while for the bulk eigenvalues of real
symmetric matrices $\mu(s) \propto s$. Wigner hypothesized \cite{Wi57}
that as with the
Poisson process, $p(0;s)$ for the bulk eigenvalues of random real symmetric
matrices could be deduced from the ansatz
\begin{equation}\label{ws1}
p(0;s) = c_1 \mu(s) \exp \Big ( - c_2 \int_0^s \mu(t) \, dt \Big )
\end{equation}
where the constants $c_1$ and $c_2$ are determined by the normalization
requirements
$$
\int_0^\infty p(0;s) \, ds = 1, \qquad
\int_0^\infty s p(0;s) \, ds = 1
$$
(the second of these says that the mean spacing is unity). Thus one arrives
at the so called Wigner surmise
\begin{equation}\label{ws}
p(0;s) = {\pi \over 2} s e^{- \pi s^2 / 4}
\end{equation}
for the spacing distribution of the bulk eigenvalues of random real symmetric
matrices.

The ansatz (\ref{ws1}) does not apply if instead of real symmetric matrices
one considers complex Hermitian matrices, or Hermitian matrices with real
quaternion elements. Examining the $2 \times 2$ case (see the introductory
article by Porter in \cite{Po65}) one sees that in the analogue of
(\ref{2.10'}), the factor $|\lambda_+-\lambda_-|$ should be replaced by
$|\lambda_+-\lambda_-|^\beta$ with $\beta=2$ (complex elements) or
$\beta=4$ (real quaternion elements). Choosing the elements to be appropriate
Gaussians, one can reclaim (\ref{ws}) and furthermore obtain
\begin{equation}\label{2.10b}
p_2(0;s) = {32 s^2 \over \pi^2} e^{-4 s^2/\pi}, \qquad
p_4(0;s) = {2^{18} s^4 \over 3^6 \pi^3} e^{-64 s^2/9 \pi}.
\end{equation}
as approximations to the spacing distributions in the cases $\beta=2$ and
$\beta=4$ respectively.

\subsection{Fredholm determinant evaluations}
A unitary invariant matrix ensemble of $N \times N$ random complex
Hermitian matrices has as its eigenvalue probability density function
\begin{equation}\label{3.1}
{1 \over C} \prod_{l=1}^N w_2(x_l) \prod_{1 \le j < k \le N}
(x_k - x_j)^2,
\end{equation}
which we will denote by UE${}_N(g)$. We know (see the lectures by Fyodorov
in this volume) that the $k$-point distribution function can be expressed
in terms of the monic orthogonal polynomials $\{p_k(x)\}_{k=0,1,\dots}$
associated with the weight function $w_2(x)$,
$$
\int_{-\infty}^\infty w_2(x) p_j(x) p_k(x) \, dx = h_j \delta_{j,k}.
$$
Thus with
\begin{eqnarray}\label{2.13'}
K_N(x,y) & = & (w_2(x) w_2(y) )^{1/2} \sum_{k=0}^{N-1} {p_k(x) p_k(y) \over h_k}
\nonumber \\
& = & (w_2(x) w_2(y) )^{1/2} {p_N(x) p_{N-1}(y) - p_N(y) p_{N-1}(x) \over x-y}
\end{eqnarray}
we have
\begin{equation}\label{3.1a}
\rho_{(k)}^{(N)}(x_1,\dots,x_k) = \det \Big [ K_N(x_j, x_l)
\Big ]_{j,l=1,\dots,k}.
\end{equation}

This structure is significant for the evaluation of the generating function
\begin{equation}\label{3.2}
E_{N,2}(J;\xi;w_2) := \Big \langle \prod_{l=1}^N(1 - \xi \chi_J^{(l)})
\Big \rangle_{{\rm UE}_N(g)}
\end{equation}
(the subscript 2 on $E_{N,2}$ indicates the exponent in (\ref{3.1})). Expanding
(\ref{3.2}) in a power series analogous to (\ref{ap2}) we obtain
\begin{equation}\label{3.2a}
E_{N,2}(J;\xi;w_2) = 1 + \sum_{k=1}^N {(-\xi)^k \over k!}
\int_J dx_1 \cdots \int_J dx_k \, \det \Big [ K_N(x_j, x_l)
\Big ]_{j,l=1,\dots,k}, 
\end{equation}
where use has been made of (\ref{3.1a}). The sum in (\ref{3.2a})
occurs in the theory of
Fredholm integral equations \cite{WW65}, 
and is in fact an expansion of the determinant
of an integral operator,
\begin{equation}
E_{N,2}(J;\xi;w_2) = \det(1 - \xi K_J)
\end{equation}
where $K_J$ is the integral operator on the interval $J$ with kernel
$K_N(x,y)$,
$$
K_N[f](x) = \int_J K_N(x,y) f(y) \, dy.
$$

It is well known that in the bulk scaling limit, independent of the precise
functional form of $w_2(x)$,
\begin{equation}\label{3.3}
\lim_{N \to \infty} a_N K_N(a_N x, a_N y) =
{\sin \pi (x-y) \over \pi (x-y) } =: K^{\rm bulk}(x,y)
\end{equation}
for a suitable scale factor $a_N$. Thus we have
\begin{equation}\label{3.4}
E_2^{\rm bulk}(J;\xi) = \det (1 - \xi K_J^{\rm bulk})
\end{equation}
where $K_J^{\rm bulk}$ is the integral operator on the interval $J$ with
kernel (\ref{3.3}) (the so called sine kernel). This is a practical formula
for the computation of $E_2^{\rm bulk}$ if we can compute the
eigenvalues $\{ \mu_j \}_{j=0,1,\dots}$ of $K_J^{\rm bulk}$, since we
have
\begin{equation}\label{3.3e}
E_2^{\rm bulk}(J;\xi) = \prod_{j=0}^\infty (1 - \xi \mu_j).
\end{equation}
In fact for $J = (-s,s)$ the eigenvalues can be computed \cite{Ga61}
by relating $K_{(-s,s)}^{\rm bulk}$ to a differential operator which
has the prolate spheroidal functions as its eigenfunctions, and using
previously computed properties of this eigensystem. 

Wigner's interest was not in complex Hermitian random matrices, but rather
real symmetric random matrices. Orthogonally invariant ensembles of the
latter have an eigenvalue probability density function of the form
\begin{equation}
{1 \over C} \prod_{l=1}^N w_1(x_l) \prod_{1 \le j < k \le N}
|x_k - x_j|,
\end{equation}
to be denoted OE${}_N(w_1)$. For such matrix ensembles, the $k$-point
distribution function can be written as a quaternion determinant (or
equivalently Pfaffian) with an underlying $2 \times 2$ matrix kernel
(see e.g.~\cite[Ch.~5]{Fo02}). From this it is possible to show that
\begin{equation}
\Big ( E_1^{\rm bulk}(J;\xi) \Big )^2 = \det ( 1 - \xi K_{1,J}^{\rm bulk} )
\end{equation}
where $K_{1,J}^{\rm bulk}$ is the integral operator on $J$ with matrix kernel
\begin{equation}
K_1^{\rm bulk}(x,y) = 
\left[
\begin{array}{cc}
\displaystyle{{\sin \pi(x-y) \over
\pi(x-y)}} & \displaystyle{{1 \over \pi } \int_0^
{ \pi  (x-y)} {\sin  t \over  t} dt} -
{1 \over 2 }{\rm sgn}(x-y)
\\[.3cm]
\displaystyle{{\partial \over \partial  x}
{\sin \pi(x-y) \over \pi
(x-y)}} & \displaystyle{{\sin \pi(x-y) \over \pi(x-y)}}
\end{array} \right].
\end{equation}
However, unlike the result (\ref{3.4}), this form has not been put to any
practical use.

Instead, as discovered by Mehta \cite{Me60}, a tractable formula results
from the scaling limit of an inter-relationship between the generating function
of an orthogonal symmetry gap probability and a unitary symmetry gap
probability. The inter-relationship states
\begin{equation}\label{me}
E_{2N,1}((-t,t);\xi;e^{-x^2/2}) \Big |_{\xi=1} =
E_{N,2}((0,t^2);\xi;y^{-1/2} e^{-y} \chi_{y > 0} ) \Big |_{\xi=1},
\end{equation}
and in the scaling limit leads to the result
\begin{equation}
E_1^{\rm bulk}((-s,s);\xi) \Big |_{\xi=1} =
\det ( 1 - K^{{\rm bulk}+}_{(-s,s)} )
\end{equation}
where $K^{{\rm bulk}+}_{(-s,s)}$ is the integral operator on $(-s,s)$
with kernel
\begin{equation}\label{2.24a}
{1 \over 2} \Big ( {\sin \pi (x-y) \over \pi (x-y) } +
{\sin \pi (x+y) \over \pi (x+y) } \Big ),
\end{equation}
which we recognize as the even part of the sine kernel (\ref{3.3}).
(For future reference we define $K^{{\rm bulk}-}_{(-s,s)}$ analogously,
except that the kernel consists of the difference of the two terms in
(\ref{2.24a}), or equivalently the odd part of the sine kernel
(\ref{3.3}).)
Because the eigenvalues $\mu_{2j}$ of the integral operator on $(-s,s)$
with kernel (\ref{3.3}) correspond to even eigenfunctions, while the
eigenvalues $\mu_{2j+1}$ correspond to odd eigenfunctions, we have that
\begin{equation}\label{ga}
E_1^{\rm bulk}((-s,s);\xi) \Big |_{\xi=1} =
\prod_{l=0}^\infty (1 - \mu_{2l}).
\end{equation}
Gaudin \cite{Ga61} used this formula, together with (\ref{pE}), to tabulate
$p_1^{\rm bulk}(0;s)$ and so test the accuracy of the Wigner surmise
(\ref{ws}). In
fact this confirmed the remarkable precision of the latter, with the
discrepancy between it and the exact value no worse than a few percent.

The case of Hermitian matrices with real quaternion elements and having a
symplectic symmetry remains. The eigenvalue p.d.f.~of the independent
eigenvalues (the spectrum is doubly degenerate) is then
\begin{equation}\label{2.26'}
{1 \over C} \prod_{l=1}^N w_4(x_l) \prod_{1 \le j < k \le N}
(x_k - x_j)^4,
\end{equation}
which we denote by SE${}_N(w_4)$. The computation of the corresponding
bulk gap probability relies on further inter-relationships between
matrix ensembles with different underlying symmetries. These apply to the
eigenvalue probability density function for Dyson's circular ensembles,
$$
{1 \over C} \prod_{1 \le j < k \le N} | e^{i \theta_k} -  e^{i \theta_j}
|^\beta,
$$
where $\beta = 1,2$ or 4 according to the underlying symmetry being
orthogonal, unitary or symplectic
respectively. The corresponding matrix ensembles are referred to as the
COE${}_N$, CUE${}_N$ and CSE${}_N$ in order.
In the $N \to \infty$ scaling limit these
ensembles correspond with the bulk of the ensembles OE${}_N(w_1)$,
UE${}_N(w_2)$ and SE${}_N(w_4)$ respectively.

The first of the required inter-relationships was formulated by Dyson
\cite{Dy62} and proved by Gunson \cite{Gu62}. It states that
\begin{equation}\label{2.1a}
{\rm alt}( {\rm COE}_N \cup {\rm COE}_N ) = {\rm CUE}_N
\end{equation}
where the operation ${\rm COE}_N \cup {\rm COE}_N$ refers to the
superposition of two independent realizations of the ${\rm COE}_N$ and
alt refers to the operation of observing only every second member of the
sequence. The second of the required inter-relationships is due to 
Dyson and Mehta \cite{DM63}. It states that
\begin{equation}\label{2.1b}
{\rm alt} \, {\rm COE}_{2N} = {\rm CSE}_N.
\end{equation}
(For generalizations of (\ref{2.1a}) and (\ref{2.1b}) to the ensembles
OE${}_N(w_1)$, UE${}_N(w_2)$ and SE${}_N(w_4)$ with particular
$w_1$, $w_2$ and $w_4$ see \cite{FR01}.)
Using (\ref{2.1a}) and (\ref{2.1b}) together one can deduce that in the
scaled limit
\begin{equation}
E_4^{\rm bulk}(0;(-s/2,s/2)) = {1 \over 2} \Big (
E_1^{\rm bulk}(0;(-s,s))  + {E_2^{\rm bulk}(0;(-s,s)) \over
E_1^{\rm bulk}(0;(-s,s)) } \Big ),
\end{equation}
which upon using (\ref{3.3e}) and (\ref{ga}) reads
\begin{equation}\label{ga1}
E_4^{\rm bulk}(0;(-s/2,s/2)) = {1 \over 2} \Big (
\prod_{l=0}^\infty(1 - \lambda_{2l}) + \prod_{l=0}^\infty(1 - \lambda_{2l+1})
\Big ).
\end{equation}
Another consequence of (\ref{2.1b}) is that
\begin{equation}\label{2.34}
p_4(0;s) = 2 p_1(1;2s).
\end{equation}
It is this relationship, used together with the approximation for
$p_4(0;s)$ in (\ref{2.10b}), which is used to approximate $p(1;s)$ as a 
smooth curve in Figure \ref{af.2}.

In summary, as a consequence of the pioneering work of Mehta, Gaudin and 
Dyson, computable formula in terms of the eigenvalues of the integral
operator on $(-s,s)$ with the sine kernel (\ref{3.3}) were obtained for
$$
E_2^{\rm bulk}((-s,s);\xi), \qquad E_1^{\rm bulk}(0;(-s,s)), \qquad
E_4^{\rm bulk}(0;(-s/2,s/2)).
$$

\section{Painlev\'e transcendent evaluations}
\setcounter{equation}{0}
\subsection{The results of Jimbo et al.}
An explicit connection between the multiple interval gap probability
$$
E_2^{\rm bulk}\Big ( \cup_{j=1}^p (a_{2j-1},a_{2j});\xi \Big )
$$
and integrable systems theory --- specifically the theory of
isomondromic deformations of linear differential equations --- was made
by Jimbo, Miwa, M\^ori and Sato in 1980. Here the endpoints
$a_1, \dots, a_{2p}$ of the gap free intervals become dynamical time like
variables, inducing flows which turn out to be integrable.

As part of this study the quantity
\begin{equation}\label{e2}
E_2^{\rm bulk}((-s,s);\xi) = \det(1 - \xi K_{(-s,s)}^{\rm bulk}) =
\prod_{j=0}^\infty (1 - \xi \mu_j)
\end{equation}
was expressed in terms of the solution of a nonlinear equation. In fact
knowledge of (\ref{e2}) is sufficient to calculate the products appearing
in (\ref{ga}) and (\ref{ga1}). Thus with
$$
D_+(s;\xi) := \prod_{j=0}^\infty(1 - \xi \mu_{2j}), \qquad
D_-(s;\xi) := \prod_{j=0}^\infty(1 - \xi \mu_{2j+1})
$$
Gaudin (see \cite{Me91}) has shown
\begin{equation}\label{y1}
\log D_{\pm}(s;\xi) = {1 \over 2} \log E_2^{\rm bulk}((-s,s);\xi) \pm
{1 \over 2} \int_0^s \sqrt{ - {d^2 \over dx^2} \log
E_2^{\rm bulk}((-x,x);\xi) } \, dx.
\end{equation}
The result of \cite{JMMS80} is that
\begin{equation}\label{jmms}
E_2^{\rm bulk}((-s,s);\xi) = \exp \int_0^{\pi s} {\sigma (u;\xi) \over u}
\, du
\end{equation}
where $\sigma(u;\xi)$ satisfies the nonlinear differential equation
\begin{equation}\label{jj1}
(u \sigma'')^2 + 4(u \sigma' - \sigma) ( u \sigma' - \sigma + (\sigma')^2 ) = 0
\end{equation}
subject to the boundary condition
$$
\sigma(u;\xi) \mathop{\sim}\limits_{u \to 0^+} - {\xi u \over \pi}.
$$

In fact the equation (\ref{jj1}) is an example of the so called $\sigma$ form
of a Painlev\'e V equation. In view of this it is appropriate to give some
background into the Painlev\'e theory, following \cite{IKSY91}. First we
remard that the
Painlev\'e differential equations are second order nonlinear equations
isolated as part of the study of Painlev\'e and his students into the
moveable singularities of the solution of such equations.
Earlier Fuchs and Poincar\'e had studied first order differential equations
of the form
\begin{equation}\label{Pp}
P(y',y,t) = 0
\end{equation}
where $P$ is a polynomial in $y', y$ with coefficients meromorphic in $t$.
In contrast to linear differential equations, nonlinear equations have the
property that the position of the  singularities of the solution will depend
in general on the initial condition. The singularities are then said to be
moveable. For example
\begin{equation}\label{Pp1}
{dy \over dt} = y^2
\end{equation}
has the general solution $y = 1/(c-t)$, where $c$ determines the 
initial condition, and so exhibits a moveable first order pole. The
nonlinear equation
$$
y {d y \over dt} = {1 \over 2}
$$
has the general solution $y = (t - c)^{1/2}$, which exhibits a moveable
branch point (essential singularity). Fuchs and Poincar\'e sought to 
classify all equations of the form (\ref{Pp}) which are free of
moveable essential singularities. They were able to show that up to an
analytic change of variables, or fractional linear transformation, the only
such equations with this property were the differential equation
of the Weierstrass ${\cal P}$-function,
\begin{equation}\label{6.12}
\Big ( {d y \over dt} \Big )^2 = 4y^3 - g_2 y - g_3,
\end{equation}
or the Riccati equation
\begin{equation}\label{6.13}
{dy \over dt} = a(t) y^2 + b(t) y + c(t)
\end{equation}
where $a, b, c$ are analytic in $t$ (note that (\ref{Pp1}) is of the
latter form).

Painlev\'e then took up the same problem as that addressed by Fuchs and
Poincar\'e, but now with respect to second order differential equations
of the form 
$$
y'' = R(y',y,t)
$$
where $R$ is a rational function in all arguments. It was found that the only
equations of this form and with no moveable essential singularities were 
either reducible to (\ref{6.12}) or (\ref{6.13}), reducible to a linear
differential equation, or were one of six new nonlinear differential
equations, now known as the Painlev\'e equations. As an explicit example
of the latter, we note the Painlev\'e V equation reads
\begin{equation}\label{PV}
y'' = \Big ( {1 \over 2y} + {1 \over 1 - y} \Big ) (y')^2
- {1 \over x} y' + {(y-1)^2 \over x^2} \Big ( \alpha y + {\beta \over
y} \Big ) + {\gamma y \over x} + {\delta y (y+1) \over y - 1}
\end{equation}
where $\alpha, \beta, \gamma$ are parameters.

An immediate question is to how (\ref{PV}) relates to (\ref{jj1}). 
 For this one must develop a Hamiltonian theory of the Painlev\'e
equations. The idea is to present a 
Hamiltonian $H=H(p,q,t;\vec{v})$, where the components of
$\vec{v}$ are parameters, such that after eliminating $p$ in the
Hamilton equations
\begin{equation}\label{6.21}
q ' = {\partial H \over \partial p}, \qquad
p' = - {\partial H \over \partial q},
\end{equation}
$q'$ and $p'$ denoting derivatives with respect to $t$,
the equation in $q$ is the appropriate Painlev\'e equation 
(in (\ref{6.21}) the role of $p$ and $q$ is interchanged relative to their
usual meaning of position and momentum in physics; here we are following
the convention of Okamoto. Malmquist \cite{Ma22} was the
first to present such Hamiltonians, although his motivation was not to
further the development of the Painlev\'e theory itself. This was left to
Okamoto in a later era, and it is aspects of his theory we will briefly
present here.

The Hamiltonian for the PV equation as presented by Okamoto \cite{OK87} is
\begin{eqnarray}
t H_V & = & q(q-1)^2p^2 - \{ (v_1 - v_2)(q-1)^2 - 2(v_1 + v_2)q(q-1) + tq
\} p \nonumber \\&& \qquad + (v_3 - v_2)(v_4 - v_2) (q-1), 
\end{eqnarray}
where the parameters are constrained by $v_1+v_2+v_3+v_4=0$ and are
further related to those in (\ref{PV}) according to
$$
\alpha = {1 \over 2}(v_3 - v_4)^2, \: \:
\beta = - {1 \over 2} (v_1 - v_2)^2, \:
\gamma = v_1 + 2 v_2 - 1, \: \: \delta = - {1 \over 2}.
$$
It turns out that, as a consequence of the Hamilton equations (\ref{6.21}),
$t H_V$ itself satisfies a nonlinear differential equation. It is this
differential equation which relates to (\ref{jj1}). Okamoto made use
of this equation for the symmetry it exhibits in the parameters
$v_1,\dots,v_4$.

The equation in question, which is fairly straightforward to derive, is
presented for the so called auxilary Hamiltonian
$$
h_V(t)  =  tH_V + (v_3 - v_2) (v_4 - v_2) - v_2 t - 2 v_2^2.
$$
Okamoto showed
$$
(th_V'')^2 - (h_V - th_V' + 2 (h_V')^2)^2 + 4
\prod_{k=1}^4(h_V'+v_k) = 0.
$$
Setting
$$
\sigma_{V}(t)  =   h_V(t) + v_2t + 2v_2^2, \qquad \nu_{j-1} = v_j - v_2
\: \:\: (j=1,\dots,4)
$$
in this one obtains the so called Jimbo-Miwa-Okamoto $\sigma$-form of the
Painlev\'e V equation
\begin{eqnarray}\label{3.12}
&&
(t \sigma_V'')^2 - \Big ( \sigma_V - t \sigma_V'
+ 2 (\sigma_V')^2 + (\nu_0 + \nu_1 + \nu_2 + \nu_3)
\sigma_V' \Big )^2 \nonumber \\
&& \quad + 4 (\nu_0 +  \sigma_V')(\nu_1 +  \sigma_V')
(\nu_2 +  \sigma_V')(\nu_3 +  \sigma_V') = 0
\end{eqnarray}
(Jimbo and Miwa \cite{JM81} arrived at (\ref{3.12}) in their study of
isomonodromic deformations of linear differential equations).
We note that (\ref{jj1}) is an example of this equation with 
$$
\nu_0 = \nu_1 = \nu_2 = \nu_3 = 0, \qquad t \mapsto - 2 i u.
$$

\subsection{Unveiling more structure}
The result of Jimbo et al.~relates to the Fredholm determinant of the
integral operator with the sine kernel. What is special about the sine
kernel that relates it to integrable systems theory? This question was
answered by Its, Izergin, Korepin
 and Slanov \cite{IIKS90} who exhibited
integrability features of all kernels of the Christoffel-Darboux
type (recall (\ref{2.13'}) in relation to the latter terminology)
\begin{equation}\label{dR3}
\xi K(x,y) = {\phi(x) \psi(y) - \phi(y) \psi(x) \over x - y},
\end{equation}
the sine kernel begin the special case 
\begin{equation}\label{sinc}
\phi(x) = \sqrt{\xi} \sin x, \qquad
\psi(y) = \sqrt{\xi} \cos y.
\end{equation}
One of their key results related to the form of the kernel $R(x,y)$
for the so called resolvent operator
$$
R_J := \xi K_J ( 1 - \xi K_J )^{-1}.
$$
With 
\begin{equation}\label{3.14'}
Q(x) := (1 - \xi K_J)^{-1} \phi, \qquad P(x) := (1 - \xi K_J)^{-1} \psi
\end{equation}
they showed
\begin{equation}\label{dR1}
R(x,y) = {Q(x) P(y) - P(x) Q(y) \over x - y}.
\end{equation}

The significance of the resolvent kernel is evident from the general formula
\begin{equation}\label{dR}
{\partial \over \partial a_j} \log \det (1 - \xi K_{(a_1,a_2)} ) =
(-1)^{j-1} R(a_j, a_j) \quad (j=1,2).
\end{equation}
To derive this formula, one notes that
\begin{eqnarray*}
\log \det (1 - \xi K_{(a_1,a_2)} ) & = & {\rm Tr} \,
\log (1 - \xi K_{(a_1,a_2)} ) \\
& = & \int_{-\infty}^\infty \log (1 - \xi K(x,x) \chi_{(a_1,a_2)}^{(x)})
\, dx.
\end{eqnarray*}
Thus
$$
{\partial \over \partial a_j} \log \det (1 - \xi K_{(a_1,a_2)} ) =
(-1)^{j-1} (1 - \xi K(a_j,a_j))^{-1} \xi K(a_j,a_j)
$$
as required.

According to (\ref{dR1})
\begin{equation}\label{dRa}
R(a_j, a_j) = - Q(x) P'(x) + P(x) Q'(x) \Big |_{x=a_j},
\end{equation}
so we see from (\ref{dR}) that the Fredholm determinant is determined by the
quantities (\ref{3.14'}) and their derivatives evaluated at the endpoints of
the interval. Indeed a close examination of the workings of 
\cite{JMMS80}, undertaken by Mehta \cite{Me91a}, Dyson \cite{Dy95} and
Tracy and Widom \cite{TW93}, revealed that 
the former study indeed proceeds via the equations (\ref{dR}) and
(\ref{dRa}), and in fact $\sigma(t)$ in (\ref{jmms}) is related to the
resolvent kernel evaluated at an endpoint by $\sigma(t) = - t
R(t/2,t/2)$. Moreover it was realized that like (\ref{dR1}) there are 
other equations contained in the working of \cite{JMMS80} which apply to all
kernels of the form (\ref{dR3}). However it was also clear that other
equations used in \cite{JMMS80} were specific to the form of
$\phi$ and $\psi$ 
in (\ref{sinc}).

Tracy and Widom were able to identify these latter properties, which are that
$\phi$ and $\psi$ are related by the coupled first order differential
equations
\begin{eqnarray}\label{7.34}
m(x) \phi'(x) & = & A(x) \phi(x) + B(x) \psi(x) \nonumber \\
m(x) \psi'(x)  & = & -C(x) \phi(x) - A(x) \psi(x)
\end{eqnarray}
where $m,A,B,C$ are polynomials. This structure allows the so called
universal equations (independent of the specific form of (\ref{dR3}))
such as (\ref{dRa}) to be supplemented by a number of case specific
equations. For some choices of $\phi$ and $\psi$ in addition to that
corresponding to sine kernel, the resulting system of equations closes.
Examples relevant to spacing distributions at the soft and hard edge of
matrix ensembles with unitary symmetry are
$$
\phi(x) = \sqrt{\xi} {\rm Ai}(x), \: \: \psi(x) = \phi'(x), \qquad \qquad
\phi(x) = \sqrt{\xi} J_a(\sqrt{x}), \: \: \psi(x) = x \phi'(x).
$$
In both these cases it was possible to obtain an evaluation of the generating
function for the corresponding gap probability in a form analogous to
(\ref{jmms}) \cite{TW94a,TW94b}.

We will make note of the hard edge result because it, by virtue of Mehta's
inter-relationship (\ref{me}), relates to the gap probability in the bulk
in the case of an underlying orthogonal symmetry. First, we define the hard
edge gap probability in the case of an underlying
unitary symmetry as the scaled limit of the ensemble (\ref{3.1}) with
$w_2(x) = x^a e^{-x} \chi_{x > 0}$. Explicitly
\begin{equation}\label{u4.1}
E_2^{\rm hard}((0,s);\xi) = \lim_{N \to \infty}
E_2\Big ( (0, {s \over 4N});\xi;x^a e^{-x} \chi_{x>0} \Big ).
\end{equation}
It was shown in \cite{Fo93a} that
\begin{equation}\label{u4.1a}
E_2^{\rm hard}((0,s);\xi) = \det (1 - \xi K^{\rm hard}_{(0,s)})
\end{equation}
where $K^{\rm hard}_{(0,s)}$ is the integral operator on $(0,s)$ with kernel
$$
K^{\rm hard}(x,y) = {J_a(x^{1/2}) y^{1/2} J_a'(y^{1/2}) - x^{1/2} 
J_a'(x^{1/2}) J_a(y^{1/2}) \over 2 ( x - y) }.
$$
As part of the study \cite{TW94b} the Fredholm determinant (\ref{u4.1a})
was given the evaluation 
\begin{equation}\label{u4.2}
E_2^{\rm hard}((0,s);\xi) = \exp \int_0^s u(t;a;\xi) {dt \over t}
\end{equation}
where $u$ satisfies the differential equation
\begin{equation}\label{6.90}
(t u'')^2 - a^2 (u')^2 - u'(4 u' + 1) (u - tu') = 0
\end{equation}
subject to the boundary condition
$$
u(t;a;\xi)
 \: \mathop{\sim}\limits_{t \to 0^+} \: - \xi t K^{\rm hard}(t,t).
$$
The equation (\ref{6.90}) is a special case of the $\sigma$ form of the
Painlev\'e III$'$ system \cite{Ok87a}.

It follows from (\ref{me}), (\ref{u4.1}) and (\ref{u4.2}) that \cite{Fo99a}
\begin{equation}\label{ch.5}
E_1^{\rm bulk}(0;(-s,s)) =
E_2^{\rm hard}(0;(0,\pi^2 s^2)) \Big |_{a=-1/2} =
\exp \int_0^{(\pi s)^2} u(t;a;\xi) \, {dt \over t}
 \Big |_{a=-1/2 \atop \xi = 1}.
\end{equation}
This is an alternative Painlev\'e transcendent evaluation to that implied
by (\ref{2.24a}), (\ref{y1}) and (\ref{jmms}). Similarly, by noting that
$$
2 \sqrt{xy} K^{\rm hard}(x^2,y^2) \Big |_{a=1/2} =
{1 \over 2} \Big ( {\sin (x - y) \over x - y} -
{\sin (x + y) \over x + y} \Big )
$$
we see from (\ref{ga1}), (\ref{u4.1a}) and (\ref{u4.2}) that \cite{Fo99a}
\begin{eqnarray}
&& E_4^{\rm bulk}(0;(-s/2,s/2)) \nonumber \\
&& \qquad = {1 \over 2} \Big (
\exp \int_0^{(\pi s)^2} u(t;a;\xi) \, {dt \over t}
 \Big |_{a=-1/2 \atop \xi = 1} +
\exp \int_0^{(\pi s)^2} u(t;a;\xi) \, {dt \over t}
 \Big |_{a=1/2 \atop \xi = 1} \Big ).
\end{eqnarray}
 
In summary, the Fredholm determinants in the expressions for the bulk gap
probabilities can each be written in terms of Painlev\'e transcendents. From
a practical viewpoint these expressions are particularly well suited for
generating power series expansions, and also allow for a numerical tabulation
of each of $E_2^{\rm bulk}(0;(-s,s))$, $E_1^{\rm bulk}(0;(-s,s))$ and
$E_4^{\rm bulk}(0;(-s,s))$, as well as $E_2^{\rm bulk}(n;(-s,s))$ for
$n \ge 1$. For the latter quantity, according to (\ref{2.5x}) we must
differentiate $E_2^{\rm bulk}((-s,s);\xi)$ with respect to $\xi$ then set
$\xi = 1$. Doing this in (\ref{jj1}) gives a coupled system of differential
equations for $\partial^j \sigma(u;\xi) / \partial \xi^j |_{\xi = 1}$
$(j=0,\dots,n)$ which is only numerically stable for small values of $n$.

\subsection{Distribution of bulk right or left nearest neighbour
spacings}
The spacing distribution refers to the distribution of the distance between
consecutive points as we move along the line left to right. Another
simple to measure statistic of this type is the distribution of the
smallest of the left neighbour spacing and right neighbour spacing for each
point. Let us denote this by $p_\beta^{\rm n.n.}(s)$ (the superscript
n.n.~stands for nearest neighbour, while the subscript $\beta$ indicates
the symmetry class). Let $E_\beta^{\rm n.n.}(0;(-s,s))$ denote the
probability that about a fixed eigenvalue at the origin, there is no
eigenvalue at distance $s$ either side. Analogous to (\ref{pE}) it is
easy to see that
\begin{equation}\label{fre0}
p_\beta^{\rm n.n.}(s) = - {d \over ds} E_\beta^{\rm n.n.}(0;(-s,s)).
\end{equation}

In the case $\beta = 2$ (unitary symmetry) the generating function
$E_\beta^{\rm n.n.}((-s,s);\xi)$ can be expressed as a Fredholm
determinant
\begin{equation}\label{fre}
E_\beta^{\rm n.n.}((-s,s);\xi) = \det ( 1 - \xi K_{(-s,s)}^{\rm n.n.})
\end{equation}
where $K_{(-s,s)}^{\rm n.n.}$ is the integral operator on $(-s,s)$ with
kernel
\begin{equation}\label{fre1}
K^{\rm n.n.}(x,y)  :=  (\pi x)^{1/2} (\pi y)^{1/2}
{\Big ( J_{a+1/2}(\pi x) J_{a-1/2}(\pi y) -
J_{a+1/2}(\pi y) J_{a-1/2}(\pi x) \Big ) \over 2(x-y)}
\end{equation}
evaluated at $a=1$. Following the strategy which leds to (\ref{u4.2}), the
Fredholm determinant (\ref{fre}) for general $a \in \zz_{\ge 0}$ can be 
characterized as the solution of a nonlinear equation. Explicitly
\cite{FO96}
\begin{equation}\label{frb}
E_\beta^{\rm n.n.}((-s,s);\xi) =
\exp \Big ( \int_0^{2 \pi s} {\sigma_a(t;\xi) \over t} \, dt \Big )
\end{equation}
where $\sigma_a$ satisfies the nonlinear equation
\begin{equation}\label{fre2}
(s \sigma_a'')^2 + 4 (-a^2 + s \sigma_a' - \sigma_a)
\Big ( (\sigma_a')^2 - \{ a - (a^2 - s \sigma_a' + \sigma_a)^{1/2} \}^2 \Big )
= 0
\end{equation}
subject to the boundary condition
$$
\sigma_a (s;\xi) \mathop{\sim}\limits_{s \to 0^+}
-\xi {2 (s/4)^{2a + 1} \over \Gamma (1/2 + a)
\Gamma(3/2 + a)}. 
$$
In the case $a=0$, (\ref{fre1}) reduces to the sine kernel and the
differential equation (\ref{fre2}) reduces to (\ref{jj1}). For general
$a$ the differential equation (\ref{fre2}) is satisfied by an auxilary
Hamiltonian for PIII (as distinct from PIII$'$) \cite{Wi03}.

\begin{figure}
\epsfxsize=12cm
\centerline{\epsfbox{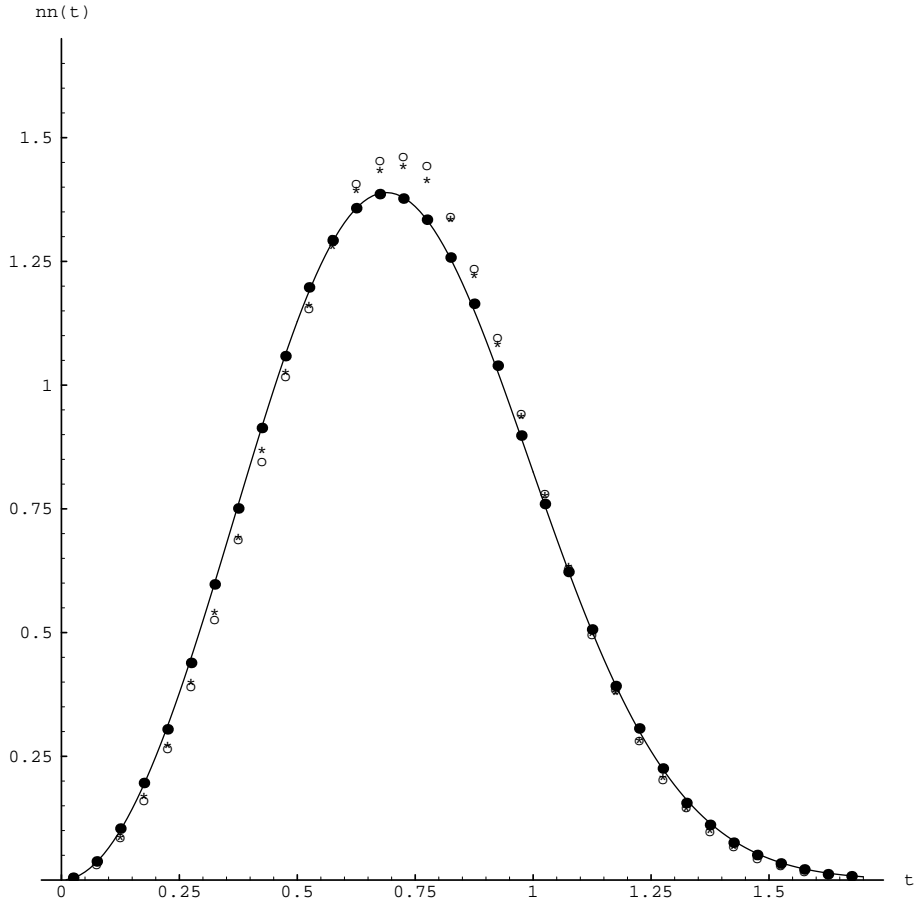}}
\caption{\label{g1} Comparison of $nn(t):= p_2^{\rm n.n.}(s)$ 
for the matrix ensembles with unitary symmetry in the bulk (continuous
curve) and for $10^6$ consecutive Riemann zeros,
starting near zero number 1 (open circles),
$10^6$ (asterisks) and $10^{20}$ (filled circles). }
\end{figure}

Substituting (\ref{frb}) in (\ref{fre0}) gives
\begin{equation}\label{fre4}
p_2^{\rm n.n.}(s) = - {\sigma_a(2 \pi s;\xi) \over 2 \pi s}
\exp \int_0^{2 \pi s} {\sigma_a(t;\xi) \over t} \, dt
\Big |_{a=\xi=1}.
\end{equation}
An application of this result can be made to the study of the zeros of the
Riemann zeta function on the critical line (Riemann zeros). We recall that
the Montgomery-Odlyzko law states that the statistics of the large
Riemann zeros coincide with the statistics of bulk eigenvalues of an
ensemble of random matrices with unitary symmetry, where both the zeros
and eigenvalues are assumed to be unfolded so as to have mean spacing
unity. As a test of this law, in \cite{FO96} the empirical determination
of $p_2^{\rm n.n.}(s)$ for large sequences of Riemann zeros, starting at
different positions along the critical line, was compared with
(\ref{fre4}). The results, which are consistent with the 
Montgomery-Odlyzko law, are reproduced in  Figure \ref{g1}. A significant
feature is that the empirical determination of $p_2^{\rm n.n}(s)$ for the
Riemann zeros is so accurate that it is not possible to compare against
an approximate form of $p_2^{\rm n.n.}(s)$ for the random matrices. Thus
the exact, readily computable,
Painlev\'e evaluation (\ref{fre4}) is of a practical importance. 

\vspace{.5cm}

\section{Gap probabilities from the Okamoto $\tau$-function theory}
\setcounter{equation}{0}
\subsection{Other stragies}
The method of Tracy and Widom may be described as being based on function
theoretic properties of Fredholm determinants. Alternative methods which
also lead to the characterization of gap probabilities in terms of the
solution of nonlinear equations have been given by a number of authors.
One alternative method is due to Adler and van Moerbeke \cite{vM01}, who
base their strategy on the fact that for suitable underlying weight
$w_2$, gap probabilities in the case of a unitary symmetry satisfy the
KP hierarchy of partial differential equations known from soliton theory.
The first member of this hierachy is then used in conjuction with a set
of equations referred to as Virasora constraints, satisfied by the gap
probabilities as a function of the endpoints of the gap free regions,
to arrive at third order equations for some single interval gap
probabilities. These third order equations are reduced to the
$\sigma$-form of the Painlev\'e theory, making use of results of
Cosgrove \cite{CS93,Co00}.
Borodin and Deift \cite{BD00} have given a method based on the
Riemann-Hilbert formulation of the resolvent kernel (\ref{dR1})
\cite{KH99}. This makes direct contact with the Schlesinger equations
from the theory of the isomonodromic deformation of linear differential
equations, and is thus closely related to the 
work of Jimbo et al.~\cite{JMMS80}.
The other approach to be mentioned is due to Forrester and Witte
\cite{FW00}. It is based on Okamoto's development of the Hamiltonian
approach to Painlev\'e systems, and proceeds by inductively constructing
sequences of multi-dimensional integral solutions of the $\sigma$ form of
the Painlev\'e equations, and identifying these solutions with gap
probabilities for certain random matrix ensembles with unitary symmetry.

For detailed accounts of all these methods, see \cite[Ch.~6\&7]{Fo02}.
In the remainder of these lectures
we will restrict ourselves to results from the work of Forrester and
Witte which relate directly to gap probabilities in the bulk.

\subsection{Direct calculation of spacing distributions}
We have taken as our objective the exact evaluation of the bulk spacing
distributions for the three symmetry classes of random matrices. So far
exact
evaluations have been presented not for the spacing distribution itself,
but rather the corresponding gap probability, which is related to the
spacing distribution by (\ref{pE}). It was realized by Forrester and
Witte \cite{FW00e} that in all three cases one of the derivatives could
be performed analytically by using theory relating to the $\sigma$ form
of the Painlev\'e transcendents.

As an explicit example, consider the result (\ref{ch.5}). It was shown in
\cite{FW00e} that
\begin{equation}\label{am.t}
{d \over ds} \exp \int_0^{(\pi s)^2} u(t;a;\xi) \, {dt \over t}
\Big |_{a=-1/2 \atop \xi = 1} = - \exp \Big ( - \int_0^{(\pi s)^2}
\tilde{u}(t) \, {dt \over t} \Big )
\end{equation}
where $\tilde{u}$ satisfies the nonlinear equation
$$
s^2 ( \tilde{u}'')^2 = (4(\tilde{u}')^2 - \tilde{u}')
(s \tilde{u}' - \tilde{u}) + {9 \over 4} ( \tilde{u}')^2 -
{3 \over 2} \tilde{u}' + {1 \over 4}
$$
subject to the boundary condition
$$
 \tilde{u}(s) \mathop{\sim}\limits_{s \to 0^+}
{s \over 3} - {s^2 \over 45} + {8 s^{5/2} \over 135 \pi}.
$$
Recalling now (\ref{pE}) we see that
\begin{equation}\label{p1b}
p_1^{\rm bulk}(0;s) = {2 \tilde{u}((\pi s/2)^2) \over s}
\exp \Big ( - \int_0^{(\pi s/2)^2} {\tilde{u}(t) \over t} \, dt \Big )
\end{equation}
(cf.~(\ref{ws})).

The identity (\ref{am.t}) can be understood from the approach to gap
probabilities of Forrester and Witte. The key advance from earlier
studies is that the generating function (\ref{ap}), with $p$ given
by (\ref{3.1}), can be generalized to the quantity
\begin{equation}\label{n.s}
E^{\rm bulk}(s;\mu;\xi) := \lim_{N \to \infty} a_N^N
\int_{-\infty}^\infty dx_1 \cdots \int_{-\infty}^\infty dx_N \,
\prod_{l=1}^N (1 - \xi \chi_{(-s/2,s/2)}^{(l)} )|s/2 - a_N x_l|^\mu
p(a_N x_1,\dots, a_N x_N)
\end{equation}
and still be characterized as the solution of a nonlinear equation. 
This is
also true at the hard and soft edges, and in the neighbourhood of a spectrum
singularity (before the generalization the latter is controlled by the
kernel (\ref{fre1})).

It is the generalization in the case of the hard edge which leads to
(\ref{am.t}). The quantity of interest is defined by
\begin{equation}\label{e2h}
E_2^{\rm hard}((0,s);\mu;\xi) = \lim_{N \to \infty}
{I_N(a) \over I_N(a+\mu)}
E_2 \Big ( (0,{s \over 4N});\xi;(x - {s \over 4N})^\mu x^a e^{-x}
\chi_{x>0} \Big )
\end{equation}
where
$$
I_N(a) := \int_0^\infty dx_1 \cdots \int_0^\infty dx_N \,
\prod_{l=1}^N e^{-x_l} x_l^a \prod_{1 \le j < k \le N} (x_k - x_j)^2
$$
(the factor $I_N(a)/I_N(a+\mu)$, which is readily evaluated in terms of
gamma functions, is chosen so that when $s=0$, (\ref{e2h}) is equal to unity).
By using theory from the Okamoto $\tau$ function approach to the
Painlev\'e systems PV and PIII$'$ it is shown in \cite{FW01a} that
$$
\tilde{E}^{\rm hard}_2((0,s);\mu;\xi) = \exp \int_0^s
u^h(t;a,\mu;\xi) \, {dt \over t},
$$
where $u^h$ satisfies the differential equation
\begin{equation}\label{uh}
(tu'')^2 - (\mu + a)^2(u')^2 - u' (4 u' + 1)(u - tu') -
{\mu(\mu+a) \over 2} u' - {\mu^2 \over 4^2} = 0.
\end{equation}
Thus we have
\begin{equation}\label{am.5}
- {1 \over \xi} {d \over ds} \exp \Big ( \int_0^s
u^h(t;a,\mu;\xi) |_{\mu=0} \, {dt \over t} \Big ) =
{s^a \over 2^{2a+2} \Gamma(a+1) \Gamma(a+2) }
\exp \Big ( \int_0^s
u^h(t;a,\mu;\xi) |_{\mu=2} \, {dt \over t} \Big ),
\end{equation}
which in the case $a=-1/2$ reduces to (\ref{am.t}).

We also read off from (\ref{am.5}) that
\begin{equation}\label{vb}
{d \over ds} \exp \int_0^{(\pi s)^2} u(t;a;\xi) {dt \over t}
\Big |_{a=1/2 \atop \xi = 1} =
- {2 \over 3} (\pi s)^2 \exp \Big ( -
\int_0^{(\pi s)^2} \tilde{v}(t) {dt \over t} \Big )
\end{equation}
where $\tilde{v}(t) = - u^h(t;a,\mu;\xi) |_{a=1/2,\mu=2,\xi=1}$ and thus
satisfies (\ref{uh}) appropriately specialized. The boundary condition
consistent with (\ref{vb}) is
\begin{equation}\label{vb1}
 \tilde{v}(t) \mathop{\sim}\limits_{t \to 0^+}
{t \over 5} ( 1 + O(t)) + {8 t^{7/2} \over 3^3 \cdot 5^3 \cdot
7 \pi} ( 1 + O(t)).
\end{equation}
Hence, according to (\ref{ga1}) and (\ref{pE}),
\begin{equation}\label{vb2}
p_4^{\rm bulk}(0;s) = 2 p_1^{\rm bulk}(0;2s) +
{2 \pi^2 s \over 3} \Big ( \tilde{v}((\pi s)^2) - 1 \Big )
\exp \Big ( - \int_0^{(\pi s)^2} \tilde{v}(t) {dt \over t} \Big ).
\end{equation}

The Okamoto $\tau$-function theory of PVI and PV allows (\ref{n.s}) to be
computed for general $\mu$, and also its generalization in which there is
a further factor $|-s/2 - a_N x_l|^a$ in the product over $l$ in the
integrand \cite{FW02}. These results allow not only the first derivative
with respect to $s$ of (\ref{jmms}) to be computed by an identity
analogous to (\ref{am.t}), but also the second derivative. In particular,
it is found that
\begin{equation}\label{p2b}
p_2^{\rm bulk}(0;s)  =  {\pi^2 \over 3} s^2
\exp \int_0^{2\pi s} v(t;\xi) \, {dt \over t}
\end{equation}
where $v$ satisfies the nonlinear equation (which can be identified in
terms of the $\sigma$ form of the PIII$'$ equation)
$$
(sv'')^2 + (v - sv') \{ v - sv' + 4 - 4 (v')^2 \}
- 16 (v')^2 = 0
$$
subject to the boundary condition
$$
v(s;\xi) \mathop{\sim}\limits_{s \to 0} 
- {1 \over 15} s^2.
$$

The exact evaluations (\ref{p1b}), (\ref{vb2}) and (\ref{p2b}) are perhaps
the most compact Painlev\'e evaluations possible for the bulk spacing
distributions. A striking feature of (\ref{p1b}) and (\ref{p2b}) is that
they are of the functional form $a(s) \exp(- \int_0^s b(t) \, dt)$ and
thus extend the Wigner surmise (\ref{ws}) and its $\beta=2$ analogue in
(\ref{2.10b}) to exact results. 

\section*{Acknowledgement}
It is a pleasure to thank the organisers for putting together such
a stimulating workshop, and program in general. Also, the financial
support of the Newton Insitute and 
the Australian Research Council is
acknowledged.


\begin{thebibliography}{10}

\bibitem{BD00}
A.~Borodin and P.~Deift.
\newblock Fredholm determinants, {Jimbo}-{Miwa}-{Ueno} tau-functions and
  representation theory.
\newblock {\em Commun. Pur. Appl. Math.}, 55:1160--1230, 2002.

\bibitem{Co00}
C.M. Cosgrove.
\newblock Chazy classes {IX-XI} of third-order differential equations.
\newblock {\em Stud. in Appl. Math.}, 104:171--228, 2000.

\bibitem{CS93}
C.M. Cosgrove and G.~Scoufis.
\newblock Painlev\'e classification of a class of differential equations of the
  second order and second degree.
\newblock {\em Stud. in Appl. Math.}, 88:25--87, 1993.

\bibitem{DE02}
I.~Dumitriu and A.~Edelman.
\newblock Matrix models for beta ensembles.
\newblock {\em J. Math. Phys.}, 43:5830--5847, 2002.

\bibitem{Dy62}
F.J. Dyson.
\newblock Statistical theory of energy levels of complex systems {III}.
\newblock {\em J. Math. Phys.}, 3:166--175, 1962.

\bibitem{Dy95}
F.J. Dyson.
\newblock The Coulomb fluid and the fifth {Painlev\'e} transcendent.
\newblock In S.-T. Yau, editor, {\em Chen Ning Yang}, page 131. International
  Press, Cambridge MA, 1995.

\bibitem{DM63}
F.J. Dyson and M.L. Mehta.
\newblock Statistical theory of the energy levels of complex systems. {IV}.
\newblock {\em J. Math. Phys.}, 4:701--712, 1963.

\bibitem{Fo02}
P.J. Forrester.
\newblock Log-gases and {Random} {Matrices}. \\
\newblock www.ms.unimelb.edu.au/\~{}matpjf/matpjf.html.

\bibitem{Fo93a}
P.J. Forrester.
\newblock The spectrum edge of random matrix ensembles.
\newblock {\em Nucl. Phys. B}, 402:709--728, 1993.

\bibitem{Fo99a}
P.J. Forrester.
\newblock Inter-relationships between gap probabilities in random matrix
  theory.
\newblock Preprint, 1999.

\bibitem{FO96}
P.J. Forrester and A.M. Odlyzko.
\newblock Gaussian unitary ensemble eigenvalues and {Riemann} $\zeta$ function
  zeros: a non-linear equation for a new statistic.
\newblock {\em Phys. Rev. E}, 54:R4493--R4495, 1996.

\bibitem{FR01}
P.J. Forrester and E.M. Rains.
\newblock Inter-relationships between orthogonal, unitary and symplectic matrix
  ensembles.
\newblock In P.M. Bleher and A.R. Its, editors, {\em Random matrix models and
  their applications}, volume~40 of {\em Mathematical Sciences Research
  Institute Publications}, pages 171--208. Cambridge University Press, United
  Kingdom, 2001.

\bibitem{FR02b}
P.J. Forrester and E.M. Rains.
\newblock Interpretations of some parameter dependent generalizations of
  classical matrix ensembles.
\newblock To appear {\em Prob. Th. Related Fields}, 2004 

\bibitem{FW00}
P.J. Forrester and N.S. Witte.
\newblock Application of the $\tau$-function theory of {Painlev\'e} equations
  to random matrices: {PIV}, {PII} and the {GUE}.
\newblock {\em Commun. Math. Phys.}, 219:357--398, 2000.

\bibitem{FW00e}
P.J. Forrester and N.S. Witte.
\newblock Exact {W}igner surmise type evaluation of the spacing distribution in
  the bulk of the scaled random matrix ensembles.
\newblock {\em Lett. Math. Phys.}, 53:195--200, 2000.

\bibitem{FW01a}
P.J. Forrester and N.S. Witte.
\newblock Application of the $\tau$-function theory of {Painlev\'e} equations
  to random matrices: {PV}, {PIII}, the {LUE}, {JUE} and {CUE}.
\newblock {\em Commun. Pure Appl. Math.}, 55:679--727, 2002.

\bibitem{FW02}
P.J. Forrester and N.S. Witte.
\newblock Application of the $\tau$-function theory of {Painlev\'e} equations
  to random matrices: {PVI}, the {JUE},{CyUE}, {cJUE} and scaled limits.
\newblock In press  {\em Nagoya J. Math.}, 2004.

\bibitem{Ga61}
M.~Gaudin.
\newblock Sur la loi limite de l'espacement des valeurs propres d'une matrice
  al\'eatoire.
\newblock {\em Nucl. Phys.}, 25:447--458, 1961.

\bibitem{Gu62}
J.~Gunson.
\newblock Proof of a conjecture of {Dyson} in the statistical theory of energy
  levels.
\newblock {\em J. Math. Phys.}, 4:752--753, 1962.

\bibitem{IIKS90}
A.R. Its, A.G. Izergin, V.E. Korepin, and N.A. Slavnov.
\newblock Differential equations for quantum correlation functions.
\newblock {\em Int. J. Mod. Phys B}, 4:1003--1037, 1990.

\bibitem{IKSY91}
K.~Iwasaki, H.~Kimura, S.~Shimomura, and M.~Yoshida.
\newblock {\em From {Gauss} to {Painlev\'e}. A modern theory of special
  functions}.
\newblock Vieweg Verlag, Braunschweig, 1991.

\bibitem{JM81}
M.~Jimbo and T.~Miwa.
\newblock Monodromony preserving deformations of linear ordinary differential
  equations with rational coefficients {II}.
\newblock {\em Physica}, 2D:407--448, 1981.

\bibitem{JMMS80}
M.~Jimbo, T.~Miwa, Y.~M\^ori, and M.~Sato.
\newblock Density matrix of an impenetrable {Bose} gas and the fifth
  {Painlev\'e} transcendent.
\newblock {\em Physica}, 1D:80--158, 1980.

\bibitem{KH99}
A.A. Kapaev and E.~Hubert.
\newblock A note on the {Lax} pairs for {Painlev\'e} equations.
\newblock {\em J. Math. Phys.}, 32:8145--8156, 1999.

\bibitem{Ma22}
J.~Malmquist.
\newblock Sur les \'equations diff\'erentialles du second ordre dont
  l'int\'egrale g\'en\'erale a ses points critiques fixes.
\newblock {\em Arkiv Mat. Astron. Fys.}, 18:1--89, 1922.

\bibitem{Me60}
M.L. Mehta.
\newblock On the statistical properties of the level-spacings in nuclear
  spectra.
\newblock {\em Nucl. Phys. B}, 18:395--419, 1960.

\bibitem{Me91a}
M.L. Mehta.
\newblock A non-linear differential equation and a {Fredholm} determinant.
\newblock {\em J. de Physique I (France)}, 2:1721--1729, 1991.

\bibitem{Me91}
M.L. Mehta.
\newblock {\em Random Matrices}.
\newblock Academic Press, New York, 2nd edition, 1991.

\bibitem{OK87}
K.~Okamoto.
\newblock Studies of the {Painlev\'e} equations. {II}. {Fifth} {Painlev\'e}
  equation {$P_{V}$}.
\newblock {\em Japan J. Math.}, 13:47--76, 1987.

\bibitem{Ok87a}
K.~Okamoto.
\newblock Studies of the {Painlev\'e} equations. {IV}. {Third} {Painlev\'e}
  equation {$P_{III}$}.
\newblock {\em Funkcialaj Ekvacioj}, 30:305--332, 1987.

\bibitem{Po65}
C.E. Porter.
\newblock {\em Statistical theories of spectra: fluctuations}.
\newblock Academic Press, New York, 1965.

\bibitem{TW93}
C.A. Tracy and H.~Widom.
\newblock Introduction to random matrices.
\newblock In G.F. Helminck, editor, {\em Geometric and quantum aspects of
  integrable systems}, volume 424 of {\em Lecture notes in physics}, pages
  407--424. Springer, New York, 1993.

\bibitem{TW94a}
C.A. Tracy and H.~Widom.
\newblock Level-spacing distributions and the {Airy} kernel.
\newblock {\em Commun. Math. Phys.}, 159:151--174, 1994.

\bibitem{TW94b}
C.A. Tracy and H.~Widom.
\newblock Level-spacing distributions and the {Bessel} kernel.
\newblock {\em Commun. Math. Phys.}, 161:289--309, 1994.

\bibitem{vM01}
P.~van Moerbeke.
\newblock Integrable lattices: random matrices and random permutations.
\newblock In P.M. Bleher and A.R. Its, editors, {\em Random matrix models and
  their applications}, volume~40 of {\em Mathematical Sciences Research
  Institute Publications}, pages 321--406. Cambridge University Press, United
  Kingdom, 2001.

\bibitem{WW65}
E.T. Whittaker and G.N. Watson.
\newblock {\em A course of modern analysis}.
\newblock CUP, Cambridge, 2nd edition, 1965.

\bibitem{Wi55}
E.P. Wigner.
\newblock Characteristic vectors of bordered matrices with infinite dimensions.
\newblock {\em Annals Math.}, 62:548--564, 1955.

\bibitem{Wi57}
E.P. Wigner.
\newblock Gatlinberg conference on neutron physics.
\newblock Oak Ridge National Laboratory Report ORNL 2309:59, 1957.

\bibitem{Wi03}
N.S. Witte.
\newblock Gap probabilities for double intervals in {H}ermitian random matrix
  ensembles as $\tau$-functions --- spectrum singularity case.
\newblock math-phy/0307063, 2003.

\end{thebibliography}

\end{document}